\documentclass[showpacs,amsmath,amssymb,preprint,aps]{revtex4}
\usepackage[T1]{fontenc}
\usepackage{amstext}
\begin{document}

\title{Travelling Waves in the Euler-Heisenberg Electrodynamics}
\author{A. D. Berm\'{u}dez Manjarres}
\email{ ad.bermudez168@uniandes.edu.co  }
\author{M. Nowakowski}
\email{mnowakos@uniandes.edu.co}
\affiliation{Departamento de Fisica\\
Universidad de los Andes\\
Cra. 1E No. 18A-10
Bogota, Colombia}
\begin{abstract}
We examine the possibility of travelling wave solutions
within the nonlinear Euler-Heisenberg electrodynamics.
Since this theory resembles in its form the electrodynamics in
matter, it is a priori not clear if there exist travelling wave solutions
with a new dispersion relation for $\omega(k)$ or if the Euler-Heisenberg
theory stringently imposes $\omega=k$ for any arbitrary ansatz
$\mathbf{E}(\xi)$ and $\mathbf{B}(\xi)$ with $\xi \equiv
\mathbf{k}\cdot\mathbf{r} -\omega t$. We show that the latter scheme
applies for the Euler-Heisenberg theory, but point out the possibility
of new solutions with $\omega \neq k$ if we go beyond the Euler-Heisenberg theory,
allowing strong fields. In case of the Euler-Heisenberg theory the quantum mechanical
effect of the travelling wave solutions remains in $\hbar$ corrections to
the energy density and the Poynting vector.
\end{abstract}
\maketitle

\section{Introduction}

In the presence of intense electromagnetic fields,
Quantum Electrodynamics predicts that the vacuum behaves like
a material medium.
This happens since starting from the one-loop level, light-light
interaction becomes possible for even number of photons.
Due to this quantum effect, the linear Maxwell theory receives
non-linear corrections. If the electromagnetic field does not change
too fast and the fields are below the so-called critical field $B{}_{c}=\frac{m_{e}^{2}}{e}$,
then the lowest order quantum corrections to classical Electrodynamics are
encoded in the Euler-Heisenberg Lagrangian \cite{1,2,3,4,5}

\begin{equation} \label{EHL}
\mathcal{L}_{EH}=a\left(\left(\mathbf{E}^{2}-\mathbf{B}^{2}\right)^{2}+7\left(\mathbf{E\cdot B}\right)^{2}\right),
\end{equation}

where
\begin{equation}
a=\frac{e^{4}}{360\pi^{2}m_{e}^{4}}.
\end{equation}

The breakdown of linearity is predicted to give rise to plenty of
new effects which do not exist in classical Electrodynamics in vacuum. At the optical
level the polarization dependent refractive index of the vacuum in
the presence of a magnetic or electric field is calculated in \cite{6}.
Calculations related to the change of the polarization of a wave due to
the birefringence of the vacuum can be found in \cite{6,7,8,9}. Other effects
include vacuum dichroism \cite{10}, second harmonic generation \cite{11,12,13,14},
parametric amplification \cite{7,15}, quantum vacuum reflection\cite{16,17},
slow light \cite{18}, photon acceleration in vacuum \cite{19}, pulse collapse  \cite{20,21}  and more (see \cite{22,23} for comprehensive reviews).
Examples of waves that are solutions to the Euler-Heisenberg  equations
but not to the classical Maxwell's equations are solitons \cite{24,25} and 
shockwaves \cite{26,27}. Both these solutions
are not travelling waves.

Worth mentioning are new developments concerning the equation of  motion
for a test body with either a charged massive particle giving rise to corrections in the
Lorentz force \cite{28}, or massless photons who now "feel" the presence of an electromagnetic
field and mimic, in a certain sense,  the motion of a massless particle in general
relativity \cite{29,30,31,32,33}. Such a self-interaction of the electromagnetic quanta or the interaction of the
photon with the field raises the question ``what is the role of a plane  wave  within such
a theory'' or, more generally, what the role of travelling waves is. Comparing the non-linear Electrodynamics
with general relativity, where plane waves as solutions exist only in the linearized version of the
theory, it is a priori not clear as to what kind of travelling waves  exist in the Euler-Heisenberg theory
and what happens to the dispersion relation. It is evident that solutions
for which the two gauge invariants $\mathbf{E}^2 -\mathbf{B}^2$ and $\mathbf{E}\cdot \mathbf{B}$ are zero,
are also solutions of the Maxwell theory with $\omega=k$. More generally, keeping $\omega=k$, the Maxwell
solution itself allows for non-zero values of the gauge invariants. The first question that we can put forward in such a context is whether these
Maxwellian solutions are also solutions in the Euler-Heisenberg theory. We will show
that the answer is affirmative if we impose a restriction. The second
question of interest is if travelling wave solutions exist in the Euler-Heisenberg theory
which have no connection to the Maxwellian case, i.e., waves with a new dispersion relation,
$\omega(k) \neq k$. We present a lengthy proof demonstrating that the only travelling wave solutions
in the Euler-Heisenberg theory are waves with $\omega(k)=k$, i.e., they are of Maxwellian type but with a restriction
on the integration constants. Interestingly, this result is not due to some physical principle which would
exclude all other solutions. From a purely mathematical point of view travelling waves exist with a new
dispersion relation, but we have to reject them on physical grounds as in these solutions the strength of the fields exceeds
the critical value allowed in the weak field approximation. We touch upon the possibility that such a restriction can, in principle,
be avoided by going beyond the Euler-Heisenberg theory. As far as the Euler-Heisenberg theory is concerned, the physical effect
of travelling wave solutions is a quantum mechanical contribution to the energy density of the waves of the Poynting vector.

The paper is organized as follows. In section 2 we review in full generality the Maxwellian case allowing for
non-zero integration constants. In section 3 we recall the salient features of the Euler-Heisenberg theory. In
section 4 we present the algebraic equations of the Euler-Heisenberg theory with
the traveling waves as an ansatz. Section 5 probes into the existence of travelling wave solutions with $\omega=k$.
In the appendix we prove that this is the only viable case. In section 6 we discuss a mathematically viable
but physically not acceptable solution with $\omega \neq k$. We present the case in order 
to argue in section 7 that
a more general Lagrangian allowing strong fields  would make a similar and analog solution possible.

\section{Maxwell's travelling waves}

The method of obtaining solutions in vacuum for the four Maxwell's
equations of classical electrodynamics is well known. It starts by
taking the Maxwell's equations, four linear first order differential
equations that involve the electric and magnetic fields, and combining
them to form two waves equations, which are second order differential equations
and then solving the wave equations. The answer is given by fields
of the form

\begin{eqnarray}
\mathbf{E} & = & \mathbf{E}(\xi)\label{eq:ansatz 1},\\
\mathbf{B} & = & \mathbf{B}(\xi)\label{eq:ansatz 2},
\end{eqnarray}
with
\begin{equation}
\xi \equiv \mathbf{k\cdot r}-\omega t.
\end{equation}
Waves with such a  dependency
on the space and time coordinates are called travelling waves.

In this paper we are interested in the travelling wave solutions
in the Euler-Heisenberg electrodynamics. In the Euler-Heisenberg case
solving the wave equation is not the most useful approach to the problem.
As a preparation for the next section and for the sake of  comparison, we present a different
way to solve the Maxwell's equation in vacuum which does not make use of the wave equation. The same approach will be
used later on to deal with the Euler-Heisenberg equations.

The magnetic Gauss's, Faraday's, electric Gauss's and Ampere-Maxwell's
laws for classical electrodynamics are

\begin{eqnarray}
\nabla\cdot\mathbf{B} & = & 0,\label{eq:magnetic gauss}\\
\nabla\times\mathbf{E} & = & -\frac{\partial\mathbf{B}}{\partial t},\label{eq:faradays}\\
\nabla\cdot\mathbf{E} & = & 0,\label{eq:electric gauss}\\
\nabla\times\mathbf{B} & = & \frac{\partial\mathbf{E}}{\partial t}.\label{eq:ampere}
\end{eqnarray}
Using a travelling wave condition as an ansatz, we can write the
Maxwell's equation as

\begin{eqnarray}
\mathbf{k}\cdot\frac{d\mathbf{B}}{d\xi} & = & 0,\\
\mathbf{k}\times\frac{d\mathbf{E}}{d\xi} & = & \omega\frac{d\mathbf{B}}{d\xi},\\
\mathbf{k}\cdot\frac{d\mathbf{E}}{d\xi} & = & 0,\\
\mathbf{k}\times\frac{d\mathbf{B}}{d\xi} & = & \omega\frac{d\mathbf{E}}{d\xi}.
\end{eqnarray}
These equations can be directly integrated to give the following algebraic
relations for the fields

\begin{eqnarray}
\mathbf{k}\cdot\mathbf{B} & = & C_{B},\\
\mathbf{B} & = & \frac{\mathbf{k}\times\mathbf{E}}{\omega}+\mathbf{d}_{B},\label{eq:algebraic F}\\
\mathbf{k}\cdot\mathbf{E} & = & C_{E},\\
\mathbf{E} & = & -\frac{\mathbf{k}\times\mathbf{B}}{\omega}+\mathbf{d}_{E}.\label{eq:algebraic A}
\end{eqnarray}
where $C_{B}$, $C_{E}$, $\mathbf{d}_{B}$ and $\mathbf{d}_{E}$
are integration constants. 

Multiplying equations (\ref{eq:algebraic F})
and (\ref{eq:algebraic A}) by $\mathbf{k}\cdot$, we see these constants
are not independent, but instead obey the relations

\begin{eqnarray}
C_{B} & = & \mathbf{k}\cdot\mathbf{d}_{B},\\
C_{E} & = & \mathbf{k}\cdot\mathbf{d}_{E}.
\end{eqnarray}
To find further relations among the quantities involved, we now replace
equation (\ref{eq:algebraic A}) into (\ref{eq:algebraic F})

\begin{equation}
\mathbf{B}=\frac{\mathbf{k}\times}{\omega}\left(-\frac{\mathbf{k}\times\mathbf{B}}{\omega}+\mathbf{d}_{E}\right)+\mathbf{d}_{B},
\end{equation}
and after some rearranging of the terms we obtain 

\begin{equation}
\mathbf{B}(1-\frac{k^{2}}{\omega^{2}})=-\frac{C_{B}}{\omega^{2}}\mathbf{k}+\mathbf{d}_{B}+\frac{\mathbf{k}\times\mathbf{d_{E}}}{\omega}.\label{eq:algebraic B}
\end{equation}
Similarly, we can replace equation (\ref{eq:algebraic F}) into equation
(\ref{eq:algebraic A}) to obtain for the electric field

\begin{equation}
\mathbf{E}(1-\frac{k^{2}}{\omega^{2}})=-\frac{C_{E}}{\omega^{2}}\mathbf{k}+\mathbf{d}_{B}-\frac{\mathbf{k}\times\mathbf{d_{B}}}{\omega}.\label{eq:algebraic E}
\end{equation}

A similar algebraic equation will emerge in the Euler-Heisenberg theory when we make
the travelling wave ansatz.

The right hand side of equations (\ref{eq:algebraic B}) and (\ref{eq:algebraic E})
are constants. Therefore the only way these equations do not lead to 
trivial constant solutions is to have the
well known dispersion relation for the classical travelling wave
$k=\omega$. In this way the equations (\ref{eq:algebraic B}) and (\ref{eq:algebraic E})
become algebraic equations that relate the constants which appear in the problem, namely

\begin{eqnarray}
\mathbf{d}_{B} & = & \frac{C_{B}}{\omega^{2}}\mathbf{k}-\frac{\mathbf{k}\times\mathbf{d_{E}}}{\omega},\\
\mathbf{d}_{E} & = & \frac{C_{E}}{\omega^{2}}\mathbf{k}+\frac{\mathbf{k}\times\mathbf{d_{B}}}{\omega}.
\end{eqnarray}

Note that if $\mathbf{d}_{B}=\mathbf{d}_{E}=0$, the equations (\ref{eq:algebraic F})
and (\ref{eq:algebraic A}) reduce to

\begin{eqnarray}
\mathbf{B} & = & \mathbf{k}\times\mathbf{E},\\
\mathbf{E} & = & -\mathbf{k}\times\mathbf{B},
\end{eqnarray}
which is the well known result that $\mathbf{k}$ and the undulatory
parts of $\mathbf{E}$ and $\mathbf{B}$ form a right handed triplet
of orthogonal vectors. This fact together with the dispersion
relations are the main results for the classical waves.

Finally, we want to find expressions for the quantities $\mathbf{E\cdot B}$
and $B^{2}-E^{2}$, which are of great importance for the generalizations
of classical electrodynamics. The first one can be obtained by direct
computation. Multiplying (13) by $\mathbf{E}$ or (15) by $\mathbf{B}$
we get
\begin{equation}
\mathbf{E\cdot B}=\mathbf{E\cdot}\mathbf{d_{B}}=\mathbf{\mathbf{d_{E}}\cdot B}.
\end{equation}
For $B^{2}-E^{2}$ we can start by squaring equation (\ref{eq:algebraic F})

\begin{eqnarray}
B^{2} & = & \left(\frac{\mathbf{k}\times\mathbf{E}}{\omega}+\mathbf{d}_{B}\right)^{2}\nonumber \\
 & = & E^{2}-\frac{C_{E}}{\omega^{2}}+d_{B}^{2}-2\mathbf{E}\cdot\left(\widehat{\mathbf{k}}\times\mathbf{d_{B}}\right)\nonumber \\
 & = & E^{2}+\frac{C_{E}}{\omega^{2}}+d_{B}^{2}-\mathbf{E\cdot d_{E}},
\end{eqnarray}
or we can square equation (\ref{eq:algebraic A}) to have

\begin{eqnarray}
E^{2} & = & B^{2}-\frac{C_{B}^{2}}{\omega^{2}}+d_{E}+2\mathbf{B}\cdot\left(\widehat{\mathbf{k}}\times\mathbf{d_{E}}\right)\nonumber \\
 & = & B^{2}+\frac{C_{B}^{2}}{\omega^{2}}+d_{E}^{2}-\mathbf{B\cdot d_{B}}.
\end{eqnarray}
With this at hand we can write $B^{2}-E^{2}$ in a few different ways

\begin{eqnarray}
B^{2}-E^{2} & = & \frac{C_{E}}{\omega^{2}}-d_{B}^{2}+2\mathbf{E}\cdot\left(\widehat{\mathbf{k}}\times\mathbf{d_{B}}\right)\nonumber \\
 & = & -\frac{C_{E}}{\omega^{2}}-d_{B}^{2}+\mathbf{E\cdot d_{E}}\nonumber \\
 & = & -\frac{C_{B}^{2}}{\omega^{2}}+d_{E}+2\mathbf{B}\cdot\left(\widehat{\mathbf{k}}\times\mathbf{d_{E}}\right)\nonumber \\
 & = & \frac{C_{B}^{2}}{\omega^{2}}+d_{E}^{2}-\mathbf{B\cdot d_{B}}.
\end{eqnarray}

As we will encounter a similar situation in the Euler-Heisenberg case, a comment on the integration constants $\mathbf{d_E}$ and $\mathbf{d_B}$ is in order. First, we mention that due to
the superposition principle in the linear Maxwell equations we can
interpret these constants as part of constant fields which then enter
the full solutions. The fact that, e.g.,  $\mathbf{d_E}$ is part of a constant
field can be seen by writing $\mathbf{B}=\mathbf{B}_0(\xi) + \mathbf{d_B}'$ and
$\mathbf{E}=\mathbf{E}_0(\xi) + \mathbf{d_E}'$. Using Faraday's law we obtain
$\mathbf{B}=\mathbf{k} \times \mathbf{E}_0 + \mathbf{d_B} + \mathbf{k} \times \mathbf{d_E}'$
where $\mathbf{d_B} + \mathbf{k} \times \mathbf{d_E}'$ is the constant magnetic field (a similar consideration can be done for the electric field).
Therefore, even if $\mathbf{k} \times \mathbf{d_E}'$ is zero, we are left with a constant magnetic contribution. 
Thus we can interpret the integration constants as parts of constant fields in which
the electromagnetic wave propagates.
Secondly, we recall
that the photon represented by $\mathbf{A}=\mathbf{\epsilon}e^{ikx}$ with $\mathbf{k} \cdot \mathbf{\epsilon}=0$ 
has two degrees of freedom with respect to $\mathbf{k}$ (two independent polarization vectors $\mathbf{\epsilon}$). 
Classically this is in correspondence with the number of parameters required to specify a plane wave in classical electrodynamics. 
Keeping the constant fields increases the number of parameters required to specify the classical field since every constant arbitrary vector has three free directions. 
This, however, does not imply that the degrees of freedom for the photon have changed as a photon which moves in a classical electromagnetic field (and every constant electromagnetic field can be considered as classical, see page 15 of \cite{34}) still has only two polarization modes \cite{7}.
 
There might exist yet another interpretation regarding the integration
constants which introduce additional degrees of freedom if we drop our
previous interpretation of  a wave in constant fields.  One such degree of freedom
could be accounted for by the breaking of the conformal symmetry at quantum level
\cite{Blaschke}.
A detailed examination of this possibility will be attempted elsewhere.

\section{Euler-Heisenberg Electrodynamics }

As in the classical electrodynamics, the Euler-Heisenberg theory consists
of four equations that determine the evolution of the electric and
the magnetic fields. The magnetic Gauss's and Faraday's
laws remain the same as in the classical case, namely

\begin{eqnarray}
\nabla\cdot\mathbf{B} & = & 0,\label{eq:EH1}\\
\nabla\times\mathbf{E} & = & -\frac{\partial\mathbf{B}}{\partial t},\label{eq:EH2}
\end{eqnarray}
These equations serve to define the electromagnetic potentials and are independent
of any Lagrangian. The second set of equations, ones that replace
the classical electric Gauss's and the Ampere-Maxwell's laws, are derived
after a variation of the Lagrangian \cite{34}. They can be written, in the
absence of electric charges and currents, as

\begin{eqnarray}
\nabla\cdot\mathbf{D} & = & 0,\label{eq:EH3}\\
\nabla\times\mathbf{H} & = & \frac{\partial\mathbf{D}}{\partial t}, \label{eq:EH4}
\end{eqnarray}
where the auxiliary fields $\mathbf{D}$ and $\mathbf{H}$ are given
by

\begin{eqnarray}
\mathbf{D} & = & \mathbf{E}+4\pi\frac{\partial\mathcal{L}_{EH}}{\partial\mathbf{E}}\nonumber \\
 & = & \mathbf{E}+\eta\left[2\mathbf{E}(E^{2}-B^{2})+7\mathbf{B}(\mathbf{E\cdot B}),\right]\label{eq:D}\\
\mathbf{H} & = & \mathbf{B}-4\pi\frac{\partial\mathcal{L}_{EH}}{\partial\mathbf{B}}\nonumber \\
 & = & \mathbf{B}+\eta\left[2\mathbf{B}(E^{2}-B^{2})-7\mathbf{E}(\mathbf{E\cdot B}),\right]\label{eq:H}
\end{eqnarray}
with

\begin{equation}
\eta=\frac{e^{4}}{45\pi m_{e}^{4}}.
\end{equation}

As is customary in classical electrodynamics, the four first order
differential equations can be combined to create two second order
wave equations \cite{25}. In this work we will not use the wave equations,
we will focus in the first order equations (31)-(34).

The symmetric gauge invariant energy-momentum tensor of this theory \cite{35,36}
is
\begin{eqnarray}
T_{\mu\nu} & = & H^{\mu\nu}F_{\nu}^{\alpha}-\mathcal{L}g_{\mu\nu},
\end{eqnarray}
where the dielectric tensor $H^{\mu\nu}$ is given by

\begin{equation}
H^{\mu\nu}=\frac{\partial\mathcal{L}}{\partial F^{\mu\nu}},
\end{equation}
and can be obtained in a simple way from $F^{\mu\nu}$ by the replacement
$E_{i}\rightarrow D_{i}$ and $B_{i}\rightarrow H_{i}$.

We follow \cite{37} and write the energy and momentum components
of the energy-momentum tensor as

\begin{eqnarray}
T^{00} & = & A\left(\frac{E^{2}+B^{2}}{8\pi}\right)+\frac{\tau}{4},\label{EH energy}\\
T^{0i} & = & A\frac{\left(\mathbf{E}\times\mathbf{B}\right)_{i}}{4\pi},\label{EH momentum}
\end{eqnarray}
where, for the weak field Euler-Heisenberg Lagrangian, the dielectric
function $A$ and the trace $\tau$ are 

\begin{eqnarray}
A & \equiv & 1+2\eta\left(E^{2}-B^{2}\right),\label{eq:dielectric f}\\
\tau & \equiv & a\left(\left(E^{2}-B^{2}\right)^{2}+7\left(\mathbf{E\cdot B}\right)^{2}\right).\label{eq:trace}
\end{eqnarray}

\section{Travelling waves in Euler-Heisenberg theory}

Our procedure is again a straightforward one, i.e., trying the ansatz
$\mathbf{E}=\mathbf{E}(\xi)$ and $\mathbf{B}=\mathbf{B}(\xi)$ into
the differential Euler-Heisenberg equations. Since the classical
dispersion relation is not a priori guaranteed to be obeyed, we look
for what conditions $\mathbf{k}$ and $\omega$ must satisfy. We can integrate the Euler-Heisenberg
equations in the same way as we did for the Maxwell's equations in section
1. We obtain

\begin{eqnarray}
\mathbf{k\cdot B} & = & C_{B},\label{eq:mgauss's EH}\\
\mathbf{B} & = & \frac{\mathbf{k}\times\mathbf{E}}{\omega}+\mathbf{d_{B}},\label{eq:faraday's EH}\\
\mathbf{k\cdot}\mathbf{D} & = & C_{D}\label{eq:egauss's EH},\\
\mathbf{D} & = & -\frac{\mathbf{k}\times\mathbf{H}}{\omega}+\mathbf{d_{D}},\label{eq:amperes's EH}
\end{eqnarray}
where $C_{B}$,$C_{D}$, $\mathbf{d}_{\mathbf{D}}$ and $\mathbf{d}_{\mathbf{B}}$ 
are constants related by taking the scalar product of (\ref{eq:faraday's EH})
and (\ref{eq:amperes's EH}) with $\mathbf{k}$: 

\begin{eqnarray}
C_{B} & = & \mathbf{k}\cdot\mathbf{d}_{B},\\
C_{D} & = & \mathbf{k}\cdot\mathbf{d}_{D}.
\end{eqnarray}

We look for the Euler-Heisenberg equivalent of equation (\ref{eq:algebraic E}).
Let us start by noticing that the auxiliary fields can be written as

\begin{eqnarray}
\mathbf{D} & = & A\mathbf{E}+7\eta(\mathbf{E}\cdot\mathbf{d_{B}})\mathbf{B},\label{eq:D2}\\
\mathbf{H} & = & A\mathbf{B}-7\eta(\mathbf{E}\cdot\mathbf{d_{B}})\mathbf{E},\label{eq:H2}
\end{eqnarray}
where $A$ is the dielectric function defined in (\ref{eq:dielectric f}).
With (\ref{eq:D2}) and (\ref{eq:H2}) the equation (\ref{eq:amperes's EH})
can be written as

\begin{equation}
A\mathbf{E}+7\eta(\mathbf{E}\cdot\mathbf{d_{B}})\mathbf{d_{B}}=-A\frac{\mathbf{k}}{\omega}\times\mathbf{B}+\mathbf{d_{D}},\label{eq:prealgebraic}
\end{equation}
where we have used (\ref{eq:faraday's EH}) to transform the terms
$7\eta(\mathbf{E}\cdot\mathbf{d_{B}})\mathbf{B}$ and $7\eta(\mathbf{E}\cdot\mathbf{d_{B}})\frac{\mathbf{k}}{\omega}\times\mathbf{E}$
into $7\eta(\mathbf{E}\cdot\mathbf{d_{B}})\mathbf{d_{B}}$. Replacing
$\mathbf{B}$ using (\ref{eq:faraday's EH}) we arrive at an algebraic equation
in which only the electric field appears

\begin{eqnarray}
A\left(1-\frac{k^{2}}{\omega^{2}}\right)\mathbf{E} & =\mathbf{d_{D}}-A\{\frac{\left(\mathbf{k\cdot E}\right)}{\omega^{2}}\mathbf{k}+\frac{\mathbf{k\times d_{B}}}{\omega}\}-7\eta\mathbf{d_{B}}\left(\mathbf{E\cdot d_{B}}\right).\label{eq:algebraic EH}
\end{eqnarray}
The dielectric function can also be put solely in terms of $\mathbf{E}$ as

\begin{equation}
A=1+2\eta\left(E^{2}\left(1-\frac{k^{2}}{\omega^{2}}\right)+\frac{\left(\mathbf{k\cdot E}\right)^{2}}{\omega^{2}}+\frac{2\mathbf{E}\cdot\left(\mathbf{k}\times\mathbf{d_{B}}\right)}{\omega}-d_{B}^{2}\right).\label{eq:dielectric E}
\end{equation}

Let us note that equation (\ref{eq:algebraic EH}) reduces to (\ref{eq:algebraic E})
in the limit $\eta\rightarrow0$, as it should be.

\section{Maxwellian case ($k=\omega$) in Euler-Heisenberg Theory}

It is well known that some solution of the Maxwell's equations are
also solutions of the Euler-Heisenberg equations \cite{6}. The simplest
examples are waves with $E^{2}-B^{2}=\mathbf{E\cdot B}=0$, where the
Euler-Heisenberg equations trivially reduce to the classical Maxwell's
ones (physically this corresponds to the fact that in QED a single free photon can propagate undisturbed \cite{40}). We shall now see that this fact can be obtained directly from (\ref{eq:algebraic EH}).
Looking for Maxwellian solutions we put $k=\omega$ into equation
(\ref{eq:algebraic EH}) to obtain

\begin{equation}
0=\mathbf{d_{D}}-A\left(\mathbf{\widehat{k}\cdot E}\right)\mathbf{\widehat{k}}-A\mathbf{\widehat{k}\times d_{B}}-7\eta\mathbf{d_{B}}\left(\mathbf{E\cdot d_{B}}\right).\label{eq:kw}
\end{equation}
Let us first assume that $\mathbf{d_{B}}$ is not parallel to
$\mathbf{\widehat{k}}$, then we can take the scalar product of
(\ref{eq:kw}) with $\mathbf{\widehat{k}}$, $\mathbf{d_{B}}$ and
$\mathbf{\widehat{k}\times d_{B}}$ (which we take as basis) to obtain the following three
equations

\begin{eqnarray}
0 & = & \mathbf{d_{D}}\cdot\mathbf{\widehat{k}}-A\left(\mathbf{\widehat{k}\cdot E}\right)-7\eta\left(\mathbf{\widehat{k}\cdot\mathbf{d_{B}}}\right)\left(\mathbf{E\cdot d_{B}}\right),\label{eq:kw1}\\
0 & = & \mathbf{d_{D}}\cdot\mathbf{d_{B}}-A\left(\mathbf{\widehat{k}\cdot E}\right)\left(\mathbf{\widehat{k}\cdot\mathbf{d_{B}}}\right)-7\eta d_{B}^{2}\left(\mathbf{E\cdot d_{B}}\right).\label{eq:kw2}\\
0 & = & \mathbf{d_{D}}\cdot\mathbf{\widehat{k}\times d_{B}}-A\left(d_{B}^{2}-\left(\mathbf{\widehat{k}\cdot\mathbf{d_{B}}}\right)^{2}\right).\cdot\label{eq:kw3}
\end{eqnarray}
From (\ref{eq:kw3}) it follows that $A=constant$. Meanwhile,
equations (\ref{eq:kw1}) and (\ref{eq:kw2}) have $\mathbf{\widehat{k}\cdot E}$
and $\mathbf{E\cdot d_{B}}$ as unknowns. Since (\ref{eq:kw1}) and
(\ref{eq:kw2}) are algebraically independent (due to our choice $\mathbf{\widehat{k}\times d_{B}}\neq0$),
we can solve $\mathbf{\widehat{k}\cdot E}$ and $\mathbf{E\cdot d_{B}}$
in terms of constants. Finally, from (\ref{eq:dielectric E}) $\mathbf{E}\cdot\left(\mathbf{k}\times\mathbf{d_{B}}\right)$
is also a constant. We have a case where there is no undulatory solution
at all.

If, on the other hand,  $\mathbf{k}$ and $\mathbf{d_{B}}$ are
parallel then equation (\ref{eq:kw}) reduces to

\begin{equation}
0=\mathbf{d_{D}}-\left(A-7\eta d_{B}^{2}\right)\left(\mathbf{\widehat{k}\cdot E}\right)\mathbf{\widehat{k}}.\label{eq:kw4}
\end{equation}

Equation (\ref{eq:kw4}) tells us that $\mathbf{d_{D}}$ has to be parallel to $\mathbf{\widehat{k}}$.
Furthermore, using (\ref{eq:dielectric E}) we can write for $A$

\begin{equation}
A=1+2\eta\left(\left(\mathbf{\widehat{k}\cdot E}\right)^{2}-d_{B}^{2}\right).\label{kwA}
\end{equation}
Then equation (\ref{eq:kw4}) together with equation (\ref{kwA})
implies that $\mathbf{\widehat{k}\cdot E}$ and $A$ are constants. This
still leave us with enough freedom for the components of $\mathbf{E}$
orthogonal to $\mathbf{\widehat{k}}$. Since $\mathbf{\widehat{k}\cdot E}$
and $A$ are constants, it can be checked that the Euler-Heisenberg equations reduces
to the Maxwell's equations. For example, the following set 
\begin{eqnarray}
\mathbf{E} & = & \mathbf{E}_{0}(\xi)+d_{E}\mathbf{\widehat{k}},\label{eq:Ek}\\
\mathbf{B} & = & \mathbf{B}_{0}(\xi)+d_{B}\mathbf{\widehat{k}},\label{eq:Bk}
\end{eqnarray}
with $\mathbf{\widehat{k}}\cdot\mathbf{E}_{0}=\mathbf{\widehat{k}}\cdot\mathbf{B}_{0}=0$
and $\mathbf{B}_{0}=\mathbf{\widehat{k}}\times\mathbf{E}_{0}$, is a
solution of both the Maxwell's and Euler-Heisenberg equations. 
Notice, however, a subtle difference. Whereas
$\mathbf{d_{B}}$ was an arbitrary constant, in the Euler-Heisenberg theory
its direction is fixed by $\mathbf{d_{B}} \propto \hat{\mathbf{k}}$.

At the end of section II we have commented on the interpretation of integration
constants in the Maxwell case.  In the Euler-Heisenberg theory constant fields
are also solutions of the corresponding equations. What we do not have here is
a general superposition principle due to the non-linearities of the equations. Interpreting the constants
in (\ref{eq:Ek}) and (\ref{eq:Bk}) as constant fields, we could say
that these equations represent a restricted superposition principle where a travelling wave and constant field can be added together 
to form a new solution if and only if  the direction of the constant field is parallel to  $\mathbf{k}$.  
An analog situation exists for two or more waves, in the sense that they can be added together 
to form a new solution to the Euler-Heisenberg equations only if they travel in the same direction \cite{40}. 
The physical interpretation given to this last effect is that the photons which travel in the same direction do not scatter from each other. 
We can then interpret (\ref{eq:Ek}) and (\ref{eq:Bk}) as a photon propagating undisturbed through a constant electromagnetic field if and only if the photon's 
motion is parallel to the direction of the background field.

Although waves (\ref{eq:Ek}) and (\ref{eq:Bk}) are also present in
the classical theory, their energy and momentum content are different in
the Euler-Heisenberg theory. For example, using (\ref{EH momentum})
we can write their momentum components as

\begin{equation}
T^{0i}=\left(1+2\eta(d_{E}^{2}-d_{B}^{2})\right)\frac{\left(\mathbf{E}\times\mathbf{B}\right)_{i}}{4\pi}.\label{T0i EH}
\end{equation}
We can see from (\ref{T0i EH}) that the photon-photon interaction
codified in the Euler-Heisenberg Lagrangian implies that the wave's
momentum density is slightly bigger when compared to the classical Poynting
vector $T_{Maxwell}^{00}=\frac{\left(\mathbf{E}\times\mathbf{B}\right)_{i}}{4\pi}$,
if $d_{E}^{2}$ is bigger than $d_{B}^{2}$ and vice versa. 

The energy density is also changed from the classical $T_{Maxwell}^{0i}=\frac{E^{2}+B^{2}}{8\pi}$
to

\begin{equation}
T^{00}=\left(1+2\eta(d_{E}^{2}-d_{B}^{2})\right)\left(\frac{E^{2}+B^{2}}{8\pi}\right)+\frac{a}{4}\left((d_{E}^{2}-d_{B}^{2})^{2}+7\left(d_{E}d_{B}\right)^{2}\right).\label{T00 EH}
\end{equation}

The new terms in the energy density and the Poynting vector proportional to 
$\eta$ and $a$ are quantum mechanical in origin.
They are 
small unless the fields become very strong, but that takes us outside
the weak field limit of the Euler-Heisenberg Lagrangian. 

In the appendix we examine all cases with $\omega \neq k$ and $ A\neq 0$ and show that they lead to trivial constant field solutions. 
The proof makes use of the fact that we can use the integration constant vectors and
$\mathbf{k}$ (or some other combinations involving cross products) as basis and decompose the electric and magnetic fields in terms of
projections in this basis.

\section{Off the light cone waves ($A=0$)}
There is a formal way to invalidate the proof presented in the appendix (this proof
demonstrates that no travelling wave solutions with $\omega \neq k$ exist in the Euler-Heisenberg theory).
Indeed it suffices to put the dielectric function $A$ to zero.  However, it is important to bring to attention
that $A=0$ is physically not viable. Indeed, such an equation would result in strong
fields violating the restriction on the theory. On the other hand, if the weak field restriction is the
only obstacle to obtain physically valid solutions, it makes sense to generalize the $A=0$ condition
to more general Lagrangians where the weak field restriction is not implemented. This seems, in principle,
possible as the Euler-Heisenberg Lagrangian (1) is a weak field version of a more general one.
As shown below, $A=0$, goes hand in hand with $\omega \neq k$, i.e., we have travelling wave
solutions off the light cone.

For these reasons it is illustrative to consider here the $A=0$ case as in the more general Lagrangian
the steps would be similar.
Taking $A=0$ in the algebraic equation (\ref{eq:algebraic EH}) gives
us the conditions

\begin{eqnarray}
1+2\eta(E^{2}-B^{2}) & = & 0,\label{eq:A1}\\
\mathbf{E\cdot B}=\mathbf{E\cdot d_{B}} & = & \beta=constant.\label{eq:A2}
\end{eqnarray}

We will call ``off light cone waves'' the waves that obey conditions (\ref{eq:A1})
and (\ref{eq:A2}).

It is easy to check that conditions (\ref{eq:A1}) and (\ref{eq:A2})
give us a solution to the full set of Euler-Heisenberg equations.
Using (\ref{eq:A1}) and (\ref{eq:A2}) the auxiliary fields become 

\begin{eqnarray}
\mathbf{D} & = & 7\eta\beta\mathbf{B},\label{eq:A3}\\
\mathbf{H} & = & -7\eta\beta\mathbf{E},\label{eq:A4}
\end{eqnarray}
and we have the strange case where the vector $\mathbf{D}$ is associated
with the magnetic field while the vector $\mathbf{H}$ is associated
with the electric field, the opposite of what one would usually expect
in electrodynamics (see, however, \cite{38}  ).

With the vectors (\ref{eq:A3}) and (\ref{eq:A4}), the modified Electric
Gauss's law (\ref{eq:EH3}) and the Ampere-Maxwell's law (\ref{eq:EH4})
become the classical magnetic Gauss's and Faraday's laws

\begin{eqnarray}
7\eta\beta\nabla\cdot\mathbf{B} & = & 0,\\
7\eta\beta\nabla\times\mathbf{E} & = & -7\eta\beta\frac{\partial\mathbf{B}}{\partial t}.
\end{eqnarray}

Notice that choosing $\beta=0$ we end up with $\mathbf{D}=\mathbf{H}=0$. Provided $A=0$, this configuration
is mathematically a solution of the Euler-Heisenberg equations.

Finally, the condition (\ref{eq:A1}) gives us an intensity dependent
dispersion relation. Indeed, using (\ref{eq:dielectric E}) we can
write

\begin{equation}
0=1+2\eta\left(E^{2}\left(1-\frac{k^{2}}{\omega^{2}}\right)+\frac{\left(\mathbf{k\cdot E}\right)^{2}}{\omega^{2}}+\frac{2\mathbf{E}\cdot\left(\mathbf{k}\times\mathbf{d_{B}}\right)}{\omega}-d_{B}^{2}\right).\label{eq:A5}
\end{equation}
As an example, consider the fields 
\begin{eqnarray}
\mathbf{E} & = & E_{0}\left(\cos\left(\xi\right)\widehat{\mathbf{x}}\text{+}\sin\left(\xi\right)\widehat{\mathbf{y}}\right),\\
\mathbf{B} & = & \frac{kE_{0}}{\omega}\left(-\sin\left(\xi\right)\widehat{\mathbf{x}}+\cos\left(\xi\right)\right)\widehat{\mathbf{y}}.
\end{eqnarray}
with $\mathbf{k}=\widehat{\mathbf{z}}$. The fields form an off light cone wave
solution to the Euler-Heisenberg equations as long as (\ref{eq:A5})
is true. Since for this example $d_{B}^{2}=\mathbf{k\cdot E}=0,$
we can calculate a dispersion relation of the form

\begin{equation}
\frac{k^{2}}{\omega^{2}}=1+\frac{1}{2\eta E_{0}^{2}}.
\end{equation}

Though unusual, the relevant energy-momentum components would simply 
read

\begin{eqnarray}
T^{00} & = & \frac{\tau}{4},\label{EH energy-1}\\
T^{0i} & = & 0.\label{EH momentum-1}
\end{eqnarray}

However, as previously stated, the off the light-cone waves are not
well-defined physical solutions. The vanishing of the dielectric function (\ref{eq:A1})
implies fields stronger than allowed by the weak field approximation
of the Euler-Heisenberg Lagrangian, i. e., 

\begin{equation}
\frac{B^{2}}{\eta}>1,
\end{equation}
whereas physically acceptable fields should range below
the critical limit $B{}_{c}=\frac{m_{e}^{2}}{e}$.

However, a more general Lagrangian, like the full Euler-Heisenberg
case, can lift this restriction.

\section{More General Lagrangian}

The Euler-Heisenberg Lagrangian (\ref{EHL}) is not the only proposed modification
to the laws of classical electrodynamics.
Indeed, we could consider the full version of the nonlinear electrodynamics arising from
quantum corrections. To avoid the problem of pair production in such a case we could
hypothetically consider an electric field below the pair production threshold and a strong magnetic field.

Let the correction to the Maxwell's Lagrangian be given by the non-linear
Lagrangian

\begin{equation}
\mathcal{L}_{NL}=\mathcal{L}_{NL}(\mathcal{F},\mathcal{G}^{2}),
\end{equation}
where the electromagnetic invariants are given by 

\begin{eqnarray}
\mathcal{F} & = & \frac{B^{2}-E^{2}}{2},\\
\mathcal{G} & = & \mathbf{E\cdot B}.
\end{eqnarray}
The pseudoscalar $\mathcal{G}$ always appears squared in the Lagrangian to preserve
the parity invariance of the theory.

In a generic form, the auxiliary fields are

\begin{eqnarray}
\mathbf{D} & = & \mathbf{E}+4\pi\frac{\partial\mathcal{L}_{NL}}{\partial\mathbf{E}}\nonumber \\
 & = & \mathbf{E}+4\pi\frac{\partial\mathcal{L}_{NL}}{\partial\mathcal{F}}\frac{\partial\mathcal{F}}{\partial\mathbf{E}}+4\pi\frac{\partial\mathcal{L}_{NL}}{\partial\mathcal{G}^{2}}\frac{\partial\mathcal{G}^{2}}{\partial\mathbf{E}}\nonumber \\
 & = & \mathbf{E}\left(1-4\pi\frac{\partial\mathcal{L}_{NL}}{\partial\mathcal{F}}\right)+8\pi\frac{\partial\mathcal{L}_{NL}}{\partial\mathcal{G}^{2}}\mathbf{B}\left(\mathbf{E\cdot B}\right),\\
\mathbf{H} & = & \mathbf{B}\left(1-4\pi\frac{\partial\mathcal{L}_{NL}}{\partial\mathcal{F}}\right)-8\pi\frac{\partial\mathcal{L}_{NL}}{\partial\mathcal{G}^{2}}\mathbf{E}\left(\mathbf{E\cdot B}\right).
\end{eqnarray}
We can again make the travelling wave ansatz and look for
solutions of the modified Maxwell equations (31) - (34).

Let us define $A \equiv 1-4\pi\frac{\partial\mathcal{L}_{NL}}{\partial\mathcal{F}}$.
Remembering that for travelling waves $\mathcal{G}=\mathbf{E\cdot B}=\mathbf{E\cdot d_{B}}$,
we can see that the conditions $A=0$ and $\mathbf{E\cdot d_{B}}=0$
guarantee vanishing auxiliary fields
\begin{equation}
\mathbf{D}=\mathbf{H}=0,
\end{equation}
and this is an immediate solution to the modified Maxwell equations. This generalizes the
situation discussed in the last section without violating the weak field restriction.
Since the full Lagrangian is given in terms of an integral, it is difficult to derive analytical expressions.
Moreover, we speculate that as in section VI,  this solution would lead to physically realizable waves with a new dispersion relation. We leave the details
to a future investigation.

We mention here that in \cite{37} the dielectric function has been calculated to all orders
for strong fields analytically up to an integral for $\mathbf{E}=0$,
$\mathbf{B} \neq 0$ and vice versa for $\mathbf{E} \neq 0$ and $\mathbf{B}=0$. 
However, if in the Maxwell Lagrangian we also set e.g. $\mathbf{E}=0$ we would not obtain travelling wave solutions and
end up with static cases. A generalization of the results in \cite{37} would be required.

\[
\]
\[
\]
$\mathbf{APPENDIX}$

In this appendix we investigate all cases of different choices of the integration constants and $\mathbf{k}$ assuming always
$\omega \neq k$. We rely on the following equations derived in the main text.
\begin{eqnarray}
A\left(1-\frac{k^{2}}{\omega^{2}}\right)\mathbf{E} & =\mathbf{d_{D}}-A\{\frac{\left(\mathbf{k\cdot E}\right)}{\omega^{2}}\mathbf{k}+\frac{\mathbf{k\times d_{B}}}{\omega}\}-7\eta\mathbf{d_{B}}\left(\mathbf{E\cdot d_{B}}\right),\label{A0}\\
A= & 1+\eta\left(E^{2}\left(1-\frac{k^{2}}{\omega^{2}}\right)+\frac{\left(\mathbf{k\cdot E}\right)^{2}}{\omega^{2}}+\frac{2\mathbf{E}\cdot\left(\mathbf{k}\times\mathbf{d_{B}}\right)}{\omega}-d_{B}^{2}\right).\label{A00}
\end{eqnarray}

\emph{Case 1:} If $\mathbf{d_{D}}\cdot\mathbf{d_{B}}=\mathbf{k}\cdot\mathbf{d_{B}}=\mathbf{k}\cdot\mathbf{d_{D}}=0$

We first analyze the case where $\mathbf{k}$, $\mathbf{d_{B}}$ and
$\mathbf{d_{D}}$ form an orthogonal basis. Multiplying (\ref{A0})
by $\mathbf{k}$, $\mathbf{d_{B}}$ and $\mathbf{d_{D}}$ we respectively
get
\begin{eqnarray}
A\left(\mathbf{k\cdot E}\right) & = & 0,\label{A1}\\
A\left(1-\frac{k^{2}}{\omega^{2}}\right) & = & -7\eta d_{B}^{2},\label{A2}\\
A\left(1-\frac{k^{2}}{\omega^{2}}\right)\left(\mathbf{E}\cdot\mathbf{d_{D}}\right) & = & d_{D}^{2}-A\mathbf{d_{D}}\cdot\left(\frac{\mathbf{k\times d_{B}}}{\omega}\right).\label{A3}
\end{eqnarray}

We see from (\ref{A2}) that $A$ is given by a constant, hence we infer from
(\ref{A1}) that $\mathbf{k\cdot E}=0$ and from (\ref{A3}) we get
that $\mathbf{E}\cdot\mathbf{d_{D}}$ is given in terms of constants.
As $\mathbf{k}$, $\mathbf{d_{B}}$ and $\mathbf{d_{D}}$ form an
orthogonal basis, $E^{2}$ can be written as

\begin{equation}
E^{2}=\left(\mathbf{E}\cdot\mathbf{\widehat{d_{B}}}\right)^{2}+\left(\mathbf{E}\cdot\mathbf{\widehat{d_{D}}}\right)^{2}\label{A4}
\end{equation}

Since $\mathbf{E}\cdot\mathbf{d_{D}}$ and $A$ are constants, when
we insert (\ref{A4}) into (\ref{A3}) we find that $\mathbf{E}\cdot\mathbf{d_{B}}$
is a constant. This case allows only trivial constants
solutions.

\emph{Case 2: $\mathbf{k}\cdot\mathbf{d_{B}}=\mathbf{k}\cdot\mathbf{d_{D}}=0$,
}$\mathbf{d_{D}}\cdot\mathbf{d_{B}}\neq0$

Taking the scalar product of (\ref{A0}) with $\mathbf{k}$, $\mathbf{d_{B}}$, $\mathbf{d_{D}}$,
$\mathbf{E}$ and $\mathbf{k\times d_{B}}$ we obtain respectively 

\begin{eqnarray}
A\left(\mathbf{k\cdot E}\right) & = & 0,\label{A5}\\
A\left(1-\frac{k^{2}}{\omega^{2}}\right)\mathbf{E}\cdot\mathbf{d_{B}} & = & \mathbf{d_{D}}\cdot\mathbf{d_{B}}-7\eta d_{B}^{2},\label{A6}\\
A\left(1-\frac{k^{2}}{\omega^{2}}\right)\left(\mathbf{E}\cdot\mathbf{d_{D}}\right) & = & d_{D}^{2}-A\mathbf{d_{D}}\cdot\left(\frac{\mathbf{k\times d_{B}}}{\omega}\right)\nonumber \\
 &  & -7\eta\mathbf{d_{D}}\cdot\mathbf{d_{B}}\left(\mathbf{E\cdot d_{B}}\right),\label{A7}\\
A\left(1-\frac{k^{2}}{\omega^{2}}\right)\mathbf{E}\cdot(\mathbf{k\times d_{B}}) & = & \mathbf{d_{D}}\cdot(\mathbf{k\times d_{B}})-A(\mathbf{k\times d_{B}})^{2}\label{A8}
\end{eqnarray}

Now we take a look at the projection. First, if $A\neq0$ then from
(\ref{A5}) $\mathbf{k\cdot E}=0.$ Since $\mathbf{d_{D}}$ is orthogonal
to $\mathbf{k}$, we can write

\begin{equation}
\mathbf{d_{D}}=a\mathbf{d_{B}}+b\left(\mathbf{k\times d_{B}}\right),\label{A9}
\end{equation}

for some constant numbers $a$ and $b$. Then, 

\begin{equation}
\mathbf{E}\cdot\mathbf{d_{D}}=a\mathbf{E}\cdot\mathbf{d_{B}}+b\mathbf{E}\cdot\left(\mathbf{k\times d_{B}}\right).\label{A10}
\end{equation}

We can insert (\ref{A10}) into (\ref{A7}) to obtain

\begin{eqnarray}
A\left(1-\frac{k^{2}}{\omega^{2}}\right)a\mathbf{E}\cdot\mathbf{d_{B}}+bA\left(1-\frac{k^{2}}{\omega^{2}}\right)\mathbf{E}\cdot\left(\mathbf{k\times d_{B}}\right) & = & d_{D}^{2}-A\mathbf{d_{D}}\cdot\left(\frac{\mathbf{k\times d_{B}}}{\omega}\right)\nonumber \\
 &-& 7\eta\mathbf{d_{D}}\cdot\mathbf{d_{B}}\left(\mathbf{E\cdot d_{B}}\right).\label{A11}
\end{eqnarray}

We can use now (\ref{A6}) and (\ref{A8}) in (\ref{A11}) to transform
its left hand side and obtain

\begin{eqnarray}
 & a\left(\mathbf{d_{D}}\cdot\mathbf{d_{B}}-7\eta d_{B}^{2}\right)+b\left(\mathbf{d_{D}}\cdot(\mathbf{k\times d_{B}})-A(\mathbf{k\times d_{B}})^{2}\right)\nonumber \\
= & d_{D}^{2}-A\mathbf{d_{D}}\cdot\left(\frac{\mathbf{k\times d_{B}}}{\omega}\right)-7\eta\mathbf{d_{D}}\cdot\mathbf{d_{B}}\left(\mathbf{E\cdot d_{B}}\right).\label{A12}
\end{eqnarray}

Our next step consists in using (\ref{A6}) to write (\ref{A2}) only
in terms of $\mathbf{d_{B}\cdot E}$. The final equations read 
\begin{eqnarray}
 & \left(\mathbf{E\cdot d_{B}}\right)\left(\mathbf{d_{D}}\cdot\mathbf{d_{B}}-7\eta d_{B}^{2}\right)\nonumber \\
 & +b\left(\left(\mathbf{E\cdot d_{B}}\right)\mathbf{d_{D}}\cdot(\mathbf{k\times d_{B}})-\frac{\left(\mathbf{d_{D}}\cdot\mathbf{d_{B}}-7\eta d_{B}^{2}\right)}{1-\frac{k^{2}}{\omega^{2}}}(\mathbf{k\times d_{B}})^{2}\right)\nonumber \\
 & =\left(\mathbf{E\cdot d_{B}}\right)d_{D}^{2}-\frac{\left(\mathbf{d_{D}}\cdot\mathbf{d_{B}}-7\eta d_{B}^{2}\right)}{1-\frac{k^{2}}{\omega^{2}}}\mathbf{d_{D}}\cdot\left(\frac{\mathbf{k\times d_{B}}}{\omega}\right)-7\eta\mathbf{d_{D}}\cdot\mathbf{d_{B}}\left(\mathbf{E\cdot d_{B}}\right)^{2}.\label{A13}
\end{eqnarray}

Equation (\ref{A13}) is a polynomial equation with constant coefficients.
Its solution gives $\mathbf{E\cdot d_{B}}$ in terms of constants.
The only way to avoid this conclusion is to have all the coefficients
of each power in $\mathbf{E\cdot d_{B}}$ to be zero individually. But it is
impossible for the coefficient of the $\left(\mathbf{E\cdot d_{B}}\right)^{2}$
to be zero
by the very same assumption we used at the beginning of this case.
\[
\]

\emph{Case 3: }If $\mathbf{d_{D}}\cdot\mathbf{d_{B}}=\mathbf{k}\cdot\mathbf{d_{B}}=0,$
and $\mathbf{k}\cdot\mathbf{d_{D}}\neq0$

Multiplying (\ref{A0}) by $\mathbf{k}$, $\mathbf{d_{B}}$ and $\mathbf{d_{D}}$
we respectively get

\begin{eqnarray}
A(\mathbf{E}\cdot\mathbf{k}) & = & \mathbf{d_{D}}\cdot\mathbf{k}\label{A14}\\
A\left(1-\frac{k^{2}}{\omega^{2}}\right) & = & -7\eta d_{B}^{2}\label{A15}\\
A\left(1-\frac{k^{2}}{\omega^{2}}\right)\left(\mathbf{E}\cdot\mathbf{d_{D}}\right) & = & d_{D}^{2}-A\left\{ \frac{(\mathbf{E}\cdot\mathbf{k})}{\omega^{2}}\mathbf{k\cdot d_{D}}-\frac{\mathbf{d_{D}}\cdot\left(\mathbf{k\times d_{B}}\right)}{\omega}\right\} \label{A16}
\end{eqnarray}

We immediately obtain from (\ref{A15}) that $A$ is a constant and we
can use this fact in (\ref{A14}) to find that $(\mathbf{E}\cdot\mathbf{k})$
is a constant. These two results together with (\ref{A16}) tell
us that $\mathbf{E}\cdot\mathbf{d_{D}}$ is a constant.

As $\mathbf{d_{B}}$ is orthogonal to $\mathbf{k}$ and $\mathbf{d_{D}}$we
can write

\begin{equation}
E^{2}=\left(\mathbf{E}\cdot\mathbf{d_{B}}\right)^{2}+F\left((\mathbf{E}\cdot\mathbf{k}),\left(\mathbf{E}\cdot\mathbf{d_{D}}\right)\right)\label{A17}
\end{equation}

where $F\left((\mathbf{E}\cdot\mathbf{k}),\left(\mathbf{E}\cdot\mathbf{d_{D}}\right)\right)$
is just a constant. We now replace (\ref{A17}) into (\ref{A00})
to arrive at an expression for $A$ 
\begin{eqnarray}
 & A=1\nonumber \\
 & +\eta\left(\left(\left(\mathbf{E}\cdot\mathbf{d_{B}}\right)^{2}+F\right)\left(1-\frac{k^{2}}{\omega^{2}}\right)+\frac{\left(\mathbf{k\cdot E}\right)^{2}}{\omega^{2}}+\frac{2\mathbf{E}\cdot\left(\mathbf{k}\times\mathbf{d_{B}}\right)}{\omega}-d_{B}^{2}\right)\label{A18}
\end{eqnarray}

The expression $\mathbf{E}\cdot\left(\mathbf{k}\times\mathbf{d_{B}}\right)^{2}$
is a constant since it can be written in terms of $(\mathbf{E}\cdot\mathbf{k}),$
and $\mathbf{E}\cdot\mathbf{d_{D}}$. Therefore using (\ref{A18})
we reach the conclusion that $\mathbf{E}\cdot\mathbf{d_{B}}$ is also a
constant.

\emph{Case 4:} If $\mathbf{d_{D}}\cdot\mathbf{d_{B}}=\mathbf{k}\cdot\mathbf{d_{D}}=0,$
and $\mathbf{k}\cdot\mathbf{d_{B}}\neq0$

First note that $\mathbf{k\times d_{B}}$ is proportional to $\mathbf{d_{D}}$.
Hence we will write $\mathbf{k\times d_{B}}=a\mathbf{d_{D}}$. 

The scalar product of (\ref{A0}) with $\mathbf{k}$, $\mathbf{d_{B}}$ and $\mathbf{d_{D}}$
gives respectively

\begin{eqnarray}
A(\mathbf{E}\cdot\mathbf{k}) & = & -7\eta\left(\mathbf{k}\cdot\mathbf{d_{B}}\right)\left(\mathbf{E\cdot d_{B}}\right),\label{A19}\\
A\left(1-\frac{k^{2}}{\omega^{2}}\right)\left(\mathbf{E}\cdot\mathbf{d_{B}}\right) & = & -A\frac{\left(\mathbf{k\cdot E}\right)}{\omega^{2}}\left(\mathbf{k}\cdot\mathbf{d_{B}}\right)-7\eta d_{B}^{2}\left(\mathbf{E\cdot d_{B}}\right),\label{A20}\\
A\left(1-\frac{k^{2}}{\omega^{2}}\right)\left(\mathbf{E}\cdot\mathbf{d_{D}}\right) & = & d_{D}^{2}-\frac{A}{\omega}ad_{D}^{2},\label{A21}\\
A\left(1-\frac{k^{2}}{\omega^{2}}\right)E^{2} & = & \mathbf{E}\cdot\mathbf{d_{D}}-A\{\frac{\left(\mathbf{k\cdot E}\right)^{2}}{\omega^{2}}\mathbf{k}+\frac{a}{\omega}\mathbf{E}\cdot\mathbf{d_{D}}\}\nonumber \\
 &  & -7\eta\left(\mathbf{E\cdot d_{B}}\right)^{2}\label{A22}
\end{eqnarray}

Replacing equation (\ref{A19}) into (\ref{A20}) leads to

\begin{equation}
A\left(1-\frac{k^{2}}{\omega^{2}}\right)=\frac{7\eta}{\omega^{2}}\left(\mathbf{k}\cdot\mathbf{d_{B}}\right)^{2}-7\eta d_{B}^{2},\label{A23}
\end{equation}

and it follows that $A$ is a constant. By virtue of (\ref{A21}) this implies that $\mathbf{E}\cdot\mathbf{d_{D}}$
is a constant.

By using equation (\ref{A00}) to write

\begin{equation}
\left(1-\frac{k^{2}}{\omega^{2}}\right)E^{2}=\frac{A-1}{\eta}-\frac{\left(\mathbf{k\cdot E}\right)^{2}}{\omega^{2}}-2\frac{a}{\omega}\mathbf{E}\cdot\mathbf{d_{D}}+d_{B}^{2}.\label{A24}
\end{equation}

and replacing (\ref{A24}) into (\ref{A22}) 

\begin{equation}
A\left(\frac{A-1}{\eta}+d_{B}^{2}-\frac{a}{\omega}\mathbf{E}\cdot\mathbf{d_{D}}\right)=\mathbf{E}\cdot\mathbf{d_{D}}-7\eta\left(\mathbf{E\cdot d_{B}}\right)^{2},\label{A25}
\end{equation}

we conclude that $\mathbf{E\cdot d_{B}}$ is a constant.

\emph{Case 5:} if $d_{D}=0$ but $d_{B}\neq0$

The scalar product of (\ref{A0}) with $\mathbf{k}$, $\mathbf{d_{B}}$ $\cdot\left(\mathbf{k\times d_{B}}\right)$
results into the following equations

\begin{eqnarray}
0 & = & A(\mathbf{k\cdot E})+7\eta\left(\mathbf{k}\cdot\mathbf{d_{B}}\right)\left(\mathbf{E\cdot d_{B}}\right),\label{A26}\\
A\left(1-\frac{k^{2}}{\omega^{2}}\right)\left(\mathbf{E}\cdot\mathbf{d_{B}}\right) & = & A\frac{(\mathbf{E}\cdot\mathbf{k})}{\omega^{2}}\mathbf{k\cdot d_{B}}-7\eta d_{B}^{2}\left(\mathbf{E\cdot d_{B}}\right),\label{A27}\\
\left(1-\frac{k^{2}}{\omega^{2}}\right)\mathbf{E}\cdot\left(\mathbf{k\times d_{B}}\right) & = & \frac{1}{\omega}\left(\mathbf{k\times d_{B}}\right)^{2}\label{A28}\\
A\left(1-\frac{k^{2}}{\omega^{2}}\right)E^{2} & = & -A\left\{ \frac{\left(\mathbf{k\cdot E}\right)^{2}}{\omega^{2}}-\frac{\mathbf{E}\cdot\left(\mathbf{k\times d_{B}}\right)}{\omega}\right\} \nonumber \\
 &  & -7\eta\left(\mathbf{E\cdot d_{B}}\right)^{2}\label{A29}
\end{eqnarray}

We can solve for $A(\mathbf{k\cdot E})$ in (\ref{A26}) and insert
it in (\ref{A27}) to obtain

\begin{equation}
A\left(1-\frac{k^{2}}{\omega^{2}}\right)=\frac{7\eta}{\omega^{2}}\left(\mathbf{k}\cdot\mathbf{d_{B}}\right)\mathbf{k\cdot d_{B}}-7\eta d_{B}^{2}.\label{A30}
\end{equation}

Again we  arrive at the conclusion that A has to be a constant. Moreover, we can read directly
from (\ref{A28}) that $\mathbf{E}\cdot\left(\mathbf{k\times d_{B}}\right)$
is a constant. From (\ref{A00}) we can write 
\begin{equation}
\left(1-\frac{k^{2}}{\omega^{2}}\right)E^{2}=\frac{A-1}{\eta}-\frac{\left(\mathbf{k\cdot E}\right)^{2}}{\omega^{2}}-2\frac{1}{\omega}\mathbf{E}\cdot\left(\mathbf{k\times d_{B}}\right)+d_{B}^{2},\label{A31}
\end{equation}
and replacing (\ref{A30}) into (\ref{A29})
\begin{equation}
-7\eta\left(\mathbf{E\cdot d_{B}}\right)^{2}=A\left[(\frac{A-1}{\eta})-\frac{1}{\omega}\mathbf{E}\cdot\left(\mathbf{k\times d_{B}}\right)+d_{B}^{2}\right].\label{A32}
\end{equation}

Independent of the numerical value of the right hand side, we easily see that $\mathbf{E\cdot d_{B}}$
is a constant.

\emph{Case 6: }If $\mathbf{d_{B}}=d_{B}\mathbf{k}$ and $\mathbf{d_{D}}\neq0.$

In this case the equation (\ref{A0}) reduces to

\begin{equation}
A\left(1-\frac{k^{2}}{\omega^{2}}\right)\mathbf{E}=\mathbf{d_{D}}-\{\frac{A}{\omega^{2}}-7\eta d_{B}^{2}\}\left(\mathbf{k\cdot E}\right)\mathbf{k}.\label{A33}
\end{equation}

We can choose $\mathbf{k}$, $\mathbf{k}\times\mathbf{d_{D}}$ and
$\mathbf{k}\times\mathbf{\left(\mathbf{k}\times\mathbf{d_{D}}\right)}$
as a basis. To make the notation more concise, let us
define $\mathbf{k}_{\perp}=\mathbf{k}\times\mathbf{\left(\mathbf{k}\times\mathbf{d_{D}}\right)}$.
It is clear from (\ref{A33}) that $\mathbf{E}$ does not have components
in the $\mathbf{k}\times\mathbf{d_{D}}$ direction, and hence $\mathbf{E}$
can be written in the following form

\begin{equation}
\mathbf{E}=\left(\mathbf{\widehat{k}\cdot E}\right)\widehat{\mathbf{k}}+\left(\mathbf{\widehat{k}_{\perp}\cdot E}\right)\widehat{\mathbf{k}}_{\perp},\label{A34}
\end{equation}
By the same token we have
\begin{equation}
\mathbf{d_{D}}=a\widehat{\mathbf{k}}+b\widehat{\mathbf{k}}_{\perp}\label{A35}
\end{equation}
for some numbers $a$ and $b$.

Equation (\ref{A34}) allows us to write

\begin{equation}
E^{2}=\left(\mathbf{\widehat{k}\cdot E}\right)^{2}+\left(\mathbf{\widehat{k}_{\perp}\cdot E}\right)^{2},\label{A36}
\end{equation}
and therefore

\begin{equation}
A=1+\eta\left(\left(\left(\mathbf{\widehat{k}\cdot E}\right)^{2}+\left(\mathbf{\widehat{k}_{\perp}\cdot E}\right)^{2}\right)\left(1-\frac{k^{2}}{\omega^{2}}\right)+\frac{\left(\mathbf{k\cdot E}\right)^{2}}{\omega^{2}}-d_{B}^{2}\right).\label{A37}
\end{equation}
The scalar product of (\ref{A33}) with $\mathbf{\widehat{k}}$, and $\mathbf{\widehat{k}}_{\perp}$
leads to the following set of equations

\begin{eqnarray}
 & \left[1+\eta\left(\left(\left(\mathbf{\widehat{k}\cdot E}\right)^{2}+\left(\mathbf{\widehat{k}_{\perp}\cdot E}\right)^{2}\right)\left(1-\frac{k^{2}}{\omega^{2}}\right)+\frac{\left(\mathbf{k\cdot E}\right)^{2}}{\omega^{2}}-d_{B}^{2}\right)\right]\left(\mathbf{\widehat{k}\cdot E}\right)\nonumber \\
 & =a-7\eta d_{B}^{2}\left(\mathbf{\widehat{k}\cdot E}\right)k^{2}\label{A38}\\
 & \left[1+\eta\left(\left(\left(\mathbf{\widehat{k}\cdot E}\right)^{2}+\left(\mathbf{\widehat{k}_{\perp}\cdot E}\right)^{2}\right)\left(1-\frac{k^{2}}{\omega^{2}}\right)+\frac{\left(\mathbf{k\cdot E}\right)^{2}}{\omega^{2}}-d_{B}^{2}\right)\right]\nonumber \\
 & \times\left(1-\frac{k^{2}}{\omega^{2}}\right)\left(\mathbf{\widehat{k}_{\perp}\cdot E}\right)\nonumber \\
 & =b & .\label{A39}
\end{eqnarray}

Equations (\ref{A38}) and (\ref{A39}) are algebraic independent
polynomials for any (non zero) value of the constants. This means that we
cannot choose any relation among $k,\, d_{B},\, a$ and $b$ to make
(\ref{A38}) proportional to (\ref{A39}). By B\'ezout's theorem \cite{39}
the systems (\ref{A38}) and (\ref{A39}) have a finite number of
solutions. These solutions will be functions of the coefficients of
the polynomials, i.e.,  of constants. Therefore we have trivial constant
solutions at hand.

On the other hand, if $\mathbf{d_{B}}$ is parallel to $\mathbf{k}$,
then equation (\ref{A31}) reduces further to

\begin{equation}
A\left(1-\frac{k^{2}}{\omega^{2}}\right)\mathbf{E}=-\{\frac{A}{\omega^{2}}\left(\mathbf{k\cdot E}\right)-7\eta d_{B}^{2}\left(\mathbf{k\cdot E}\right)-d_{D}\}\mathbf{k}.\label{A40}
\end{equation}

There are two ways to solve equation (\ref{A40}). The first is letting $k=\omega$
that leads to the condition $\mathbf{k\cdot E}=constant$ which is
identical to the classical Gauss law and also leads to a classical
solution to the Maxwell's equations. The other solution is to set
$A=0$ which also leads to $\mathbf{k\cdot E}=constant$, but we know
from section 6 that this kind of waves are not viable solutions.
\[
\]

\emph{Case 7: }$\mathbf{d_{D}}=d_{D}\mathbf{k}$ and $\mathbf{d_{B}}\neq0.$

For this case, equation (\ref{A0}) reduces to

\begin{equation}
A\left(1-\frac{k^{2}}{\omega^{2}}\right)\mathbf{E}=d_{D}\mathbf{k}-A\{\frac{\left(\mathbf{k\cdot E}\right)}{\omega^{2}}\mathbf{k}+\frac{\mathbf{k\times d_{B}}}{\omega}\}-7\eta\mathbf{d_{B}}\left(\mathbf{E\cdot d_{B}}\right).\label{A41}
\end{equation}

By taking the dot product with $\mathbf{k\times d_{B}}$ we get

\begin{equation}
\mathbf{\mathbf{E}\cdot\left(k\times d_{B}\right)}=C=constant.\label{A42}
\end{equation}

Similar to the previous case, if $\mathbf{d_{B}}$ is not parallel
to $\mathbf{k}$, then we can choose as a basis the vectors $\mathbf{k}$,
$\mathbf{k}\times\mathbf{d_{B}}$ and, $\mathbf{\widehat{k}_{\perp}}$
where $\mathbf{\widehat{k}_{\perp}}=\mathbf{k}\times\mathbf{\left(\mathbf{k}\times\mathbf{d_{B}}\right)}$.
In this way we can write $\mathbf{E}=\left(\mathbf{E\cdot\mathbf{\widehat{k}}}\right)\mathbf{\widehat{k}}+\left(\mathbf{E}\cdot\mathbf{\widehat{k}_{\perp}}\right)\mathbf{\widehat{k}_{\perp}}$,
and therefore 
\begin{eqnarray}
E^{2} & = & \left(\mathbf{\widehat{k}\cdot E}\right)^{2}+\left(\mathbf{\widehat{k}_{\perp}\cdot E}\right)^{2}+C,\label{A43}\\
A & = & 1\nonumber \\
 &  & +\eta[\left(\left(\mathbf{\widehat{k}\cdot E}\right)^{2}+\left(\mathbf{\widehat{k}_{\perp}\cdot E}\right)^{2}+C\right)\label{A44}\\
 &  & \times(\left(1-\frac{k^{2}}{\omega^{2}}\right)+\frac{\left(\mathbf{k\cdot E}\right)^{2}}{\omega^{2}}-d_{B}^{2})],\\
\mathbf{E\cdot d_{B}} & = & \left(\mathbf{E\cdot\mathbf{\widehat{k}}}\right)\mathbf{\widehat{k}\cdot d_{B}}+\left(\mathbf{E}\cdot\mathbf{\widehat{k}_{\perp}}\right)\mathbf{\widehat{k}_{\perp}\cdot d_{B}}.\label{A45}
\end{eqnarray}

We can then write the equations for the projections in $\mathbf{\widehat{k}}$,
and $\mathbf{\widehat{k}}_{\perp}$ to get

\begin{eqnarray}
 & \left[1+\eta\left(\left(\left(\mathbf{\widehat{k}\cdot E}\right)^{2}+\left(\mathbf{\widehat{k}_{\perp}\cdot E}\right)^{2}+C^{2}\right)\left(1-\frac{k^{2}}{\omega^{2}}\right)
+\frac{\left(\mathbf{k\cdot E}\right)^{2}}{\omega^{2}}-d_{B}^{2}\right)\right]\left(\mathbf{\widehat{k}\cdot E}\right)\nonumber \\
 & =d_{D}k-7\eta\left(\mathbf{\mathbf{\widehat{k}\cdot d_{B}}}\right)\left(\left(\mathbf{E\cdot\mathbf{\widehat{k}}}\right)\mathbf{\widehat{k}\cdot d_{B}}
+\left(\mathbf{E}\cdot\mathbf{\widehat{k}_{\perp}}\right)\mathbf{\widehat{k}_{\perp}\cdot d_{B}}\right)\label{A46}\\
 & \left[1+\eta\left(\left(\left(\mathbf{\widehat{k}\cdot E}\right)^{2}+\left(\mathbf{\widehat{k}_{\perp}\cdot E}\right)^{2}+C^{2}\right)\left(1-\frac{k^{2}}{\omega^{2}}\right)+\frac{\left(\mathbf{k\cdot E}\right)^{2}}{\omega^{2}}-d_{B}^{2}\right)\right]\left(1-\frac{k^{2}}{\omega^{2}}\right)\left(\mathbf{\widehat{k}_{\perp}\cdot E}\right)\nonumber \\
 & =7\eta\left(\mathbf{\mathbf{\widehat{k}_{\perp}\cdot d_{B}}}\right)\left(\left(\mathbf{E\cdot\mathbf{\widehat{k}}}\right)\mathbf{\widehat{k}\cdot d_{B}}+\left(\mathbf{E}\cdot\mathbf{\widehat{k}_{\perp}}\right)\mathbf{\widehat{k}_{\perp}\cdot d_{B}}\right) & .\label{A47}
\end{eqnarray}

As in the previous case, equations (\ref{A46}) and (\ref{A47}) are
algebraically independent, and therefore only admit a finite number
of constant solutions.

For $\mathbf{d_{B}}$ parallel to $\mathbf{k}$ we can write (\ref{A41})
as

\begin{equation}
A\left(1-\frac{k^{2}}{\omega^{2}}\right)\mathbf{E}=d_{D}\mathbf{k}-A\frac{\left(\mathbf{k\cdot E}\right)}{\omega^{2}}\mathbf{k}-7\eta d_{B}^{2}\left(\mathbf{E\cdot k}\right)\mathbf{k},\label{A48}
\end{equation}

but $\mathbf{E}=\frac{\left(\mathbf{k\cdot E}\right)}{k^{2}}\mathbf{k}$
and $A=1+\eta\left(E^{2}\left(1-\frac{k^{2}}{\omega^{2}}\right)+\frac{\left(\mathbf{k\cdot E}\right)^{2}}{\omega^{2}}-d_{B}^{2}\right)=1+\eta\left(\frac{\left(\mathbf{k\cdot E}\right)^{2}}{k^{2}}-d_{B}^{2}\right)$
and therefore we can write

\begin{equation}
\left(1+\eta\left(\frac{\left(\mathbf{k\cdot E}\right)^{2}}{k^{2}}-d_{B}^{2}\right)\right)\frac{\left(\mathbf{k\cdot E}\right)}{k}=d_{D}k-7\eta d_{B}^{2}k\left(\mathbf{E\cdot k}\right),\label{A49}
\end{equation}
which is an algebraic equation for $\left(\mathbf{k\cdot E}\right)$
in terms of constant coefficients and therefore we again haave a trivial
constant solution for the fields.
\[
\]

\emph{Case 8: } $\mathbf{k}$, $\mathbf{d_{B}}$, $\mathbf{d_{D}}$
are parallel.

This case is trivial. When $\mathbf{k}$, $\mathbf{d_{B}}$, $\mathbf{d_{D}}$
are parallel and neither $A$ nor $1-\frac{k^{2}}{\omega^{2}}$ vanish,
then we can write (\ref{A0}) as

\begin{equation}
A\left(1-\frac{k^{2}}{\omega^{2}}\right)\mathbf{E}=d_{D}\mathbf{k}-A\frac{\left(\mathbf{k\cdot E}\right)}{\omega^{2}}\mathbf{k}-7\eta d_{B}^{2}\left(\mathbf{E\cdot k}\right)\mathbf{k}.\label{A50}
\end{equation}

But $\mathbf{E}=\frac{\left(\mathbf{k\cdot E}\right)}{k^{2}}\mathbf{k}$
and $A=1+\eta\left(E^{2}\left(1-\frac{k^{2}}{\omega^{2}}\right)+\frac{\left(\mathbf{k\cdot E}\right)^{2}}{\omega^{2}}-d_{B}^{2}\right)=1+\eta\left(\frac{\left(\mathbf{k\cdot E}\right)^{2}}{k^{2}}-d_{B}^{2}\right)$
and therefore we can write

\begin{equation}
\left(1+\eta\left(\frac{\left(\mathbf{k\cdot E}\right)^{2}}{k^{2}}-d_{B}^{2}\right)\right)\frac{\left(\mathbf{k\cdot E}\right)}{k}=d_{D}k-7\eta d_{B}^{2}k\left(\mathbf{E\cdot k}\right),\label{A51}
\end{equation}
which is an algebraic equation for $\left(\mathbf{k\cdot E}\right)$
in terms of constant coefficients and therefore we again have a trivial
constant solution for the fields.
\[
\]

\emph{Case 9}: None of \emph{ }$\mathbf{k}$, $\mathbf{d_{B}}$
and $\mathbf{d_{D}}$ are parallel or orthogonal to any of the others.

Taking the scalar product of (\ref{A0}) with $\mathbf{k}$, $\mathbf{d_{B}}$, $\mathbf{k\times d_{B}}$
and $\mathbf{E}$ we respectively get

\begin{eqnarray}
A(\mathbf{E}\cdot\mathbf{k}) & = & \mathbf{d_{D}\cdot k}-7\eta\left(\mathbf{k}\cdot\mathbf{d_{B}}\right)\left(\mathbf{E\cdot d_{B}}\right),\label{A53}\\
A\left(1-\frac{k^{2}}{\omega^{2}}\right)\left(\mathbf{E}\cdot\mathbf{d_{B}}\right) & = & \mathbf{d_{D}\cdot\mathbf{d_{B}}}+A\frac{(\mathbf{E}\cdot\mathbf{k})}{\omega^{2}}\mathbf{k\cdot d_{B}}\nonumber \\
 &  & -7\eta d_{B}^{2}\left(\mathbf{E\cdot d_{B}}\right),\label{A54}\\
A\left(1-\frac{k^{2}}{\omega^{2}}\right)\mathbf{E}\cdot\left(\mathbf{k\times d_{B}}\right) & = & \mathbf{d_{D}\cdot k\times d_{B}}+\frac{A}{\omega}\left(\mathbf{k\times d_{B}}\right)^{2}.\label{A55}
\end{eqnarray}

As $\mathbf{k}$, $\mathbf{d_{B}}$ and $\mathbf{k\times d_{B}}$
are not parallel they form a basis and we can write any other vector,
like $\mathbf{E}$ and $\mathbf{d_{D}}$, as a linear combination
of them. This means that $E^{2}$ (and therefore $A$) can be written
in terms of $\mathbf{E}\cdot\mathbf{k}$, $\mathbf{E}\cdot\mathbf{d_{B}}$
and $\mathbf{E}\cdot\left(\mathbf{k\times d_{B}}\right)$. Moreover,
$E^{2}$ (and therefore $A$) will contain a term $\left(\mathbf{E}\cdot\widehat{\left(\mathbf{k\times d_{B}}\right)}\right)^{2}$,
and therefore equation (\ref{A55}) will have a term $\left(\mathbf{E}\cdot\widehat{\left(\mathbf{k\times d_{B}}\right)}\right)^{3}$.
This cubic term cannot be eliminated by any choice of the constants,
and therefore equation (\ref{A55}) cannot be reduced to equation
(\ref{A53}) or (\ref{A54}). Using the same argument, equations
(\ref{A54}) will have a cubic term of the form $\left(\mathbf{E}\cdot\mathbf{d_{B}}\right)^{3}$
that cannot be eliminated and therefore equation (\ref{A54}) cannot
be reduced to equation (\ref{A53}). We have then a system of three
algebraically independent equations for the three unknowns. We can
use B\'ezout's theorem to say that the system allows only for a
finite number of solutions that will be given in terms of constants.
Therefore, this case also leads to a trivial constant solution.

This completes our proof that all $\omega \neq k$ cases lead to trivial constant solutions
assuming $A \neq 0$.
\begin{acknowledgments}
We thank the Faculty of Science at the Universidad de los Andes
and the administrative department of science, technology and
innovation of Colombia (Colciencias) for financial support.
\end{acknowledgments}

\end{document}